\documentclass[preprint2]{aastex}

\newcommand{\fuvcenter}{1530~\AA}
\newcommand{\nuvcenter}{2310~\AA}

\newcommand{\FUV}{{\it FUV }}
\newcommand{\NUV}{{\it NUV }}
\newcommand{\galex}{\textit{GALEX }}
\newcommand{\type}{\texttt{type}}

\shorttitle{Clustering Properties of rest-frame UV selected galaxies. I}

\shortauthors{...}
\usepackage{natbib}
\bibpunct{(}{)}{;}{a}{}{,}
\newdimen\hssize
\begin{document}

\title{Clustering Properties of restframe UV selected galaxies I:\\
the correlation length derived from GALEX data in the local Universe}

\author{ 
  Bruno Milliard\altaffilmark{1}, 
  S\'ebastien Heinis\altaffilmark{1,2},
  J\'er\'emy Blaizot\altaffilmark{1,3}, 
  St\'ephane Arnouts\altaffilmark{1},\\
  David Schiminovich\altaffilmark{4}, 
  Tam\'as Budav\'ari\altaffilmark{2},
  Jos\'e Donas\altaffilmark{1}, 
  Marie Treyer\altaffilmark{1}, 
  Michel Laget\altaffilmark{1},
  Maurice Viton\altaffilmark{1},
  Ted K. Wyder\altaffilmark{5},
  Alex S. Szalay\altaffilmark{2},
  Tom A. Barlow\altaffilmark{5}, 
  Karl Forster\altaffilmark{5},\\
  Peter G. Friedman\altaffilmark{5},
  D. Christopher Martin\altaffilmark{5},
  Patrick Morrissey\altaffilmark{5},
  Susan G. Neff\altaffilmark{6},\\
  Mark Seibert\altaffilmark{5},
  Todd Small\altaffilmark{5},
  Luciana Bianchi\altaffilmark{7},
  Timothy M. Heckman\altaffilmark{2},
  Young-Wook Lee\altaffilmark{8},
  Barry F. Madore\altaffilmark{9,10},
  R. Michael Rich\altaffilmark{11},
  Barry Y. Welsh\altaffilmark{12},
  Sukyoung K. Yi\altaffilmark{8}
  and C. K. Xu\altaffilmark{5}}

\altaffiltext{1}{Laboratoire d'Astrophysique de Marseille, BP 8, Traverse
  du Siphon, 13376 Marseille Cedex 12, France}

\altaffiltext{2}{Department of Physics and Astronomy, The Johns Hopkins
  University, Homewood Campus, Baltimore, MD 21218}

\altaffiltext{3}{Max Planck Institut f\"ur astrophysik, D-85748
  Garching, Germany}

\altaffiltext{4}{Department of Astronomy, Columbia University, New
York, NY 10027}

\altaffiltext{5}{California Institute of Technology, MC 405-47, 1200 East
  California Boulevard, Pasadena, CA 91125}

\altaffiltext{6}{Laboratory for Astronomy and Solar Physics, NASA Goddard
  Space Flight Center, Greenbelt, MD 20771}

\altaffiltext{7}{Center for Astrophysical Sciences, The Johns Hopkins
  University, 3400 N. Charles St., Baltimore, MD 21218}

\altaffiltext{8}{Center for Space Astrophysics, Yonsei University, Seoul
  120-749, Korea}

\altaffiltext{9}{Observatories of the Carnegie Institution of Washington,
  813 Santa Barbara St., Pasadena, CA 91101}

\altaffiltext{10}{NASA/IPAC Extragalactic Database, California Institute
  of Technology, Mail Code 100-22, 770 S. Wilson Ave., Pasadena, CA 91125}

\altaffiltext{11}{Department of Physics and Astronomy, University of
  California, Los Angeles, CA 90095}

\altaffiltext{12}{Space Sciences Laboratory, University of California at
  Berkeley, 601 Campbell Hall, Berkeley, CA 94720}

\begin{abstract}
  We present the first measurements of the angular correlation
  function of galaxies selected in the far (\fuvcenter) and near
  (\nuvcenter) Ultraviolet from the \galex survey fields overlapping SDSS
  DR5 in low galactic extinction regions. The area used covers $120$
  sqdeg (\galex - MIS) down to magnitude AB $=22$, yielding a total of
  100,000 galaxies. The mean correlation length is $\sim3.7 \pm
  0.6$~Mpc and no significant trend is seen for this value as a
  function of the limiting apparent magnitude or between the \galex
  bands. This estimate is close to that found from samples of blue
  galaxies in the local universe selected in the visible, and similar
  to that derived at $z\simeq3$ for LBGs with similar rest frame
  selection criteria. This result supports models that predict
  anti-biasing of star forming galaxies at low redshift, and brings an
  additional clue to the downsizing of star formation at $z<1$.
\end{abstract}

\keywords{Galaxies: UV - Correlation Function Evolution - Star
Formation}

\section{Introduction}

In the current paradigm of structure formation, the bulk of the most
massive systems form in a cold dark matter-dominated universe by the
merging of less massive units formed earlier. In parallel to this
hierarchical evolution, recent observations point to the so-called
``downsizing'', namely the fact that in galaxies having high baryonic
masses the bulk of stars formed at high redshift ($z \gtrsim 1$),
while in galaxies having low baryonic masses the bulk of stars formed
at lower redshift \citep[][and also \citet{DeLucia_2006} and
\citet{Neistein_2006} for results from simulations]{Cowie_1996,
Heavens_2004, Bundy_2006, Jimenez_2005}. The star formation efficiency
shows a strong decline at $0<z<1$, as measured by the evolution of the
star formation rate density \citep{Hopkins_2006, Lilly_1996,
Schiminovich_2005, Sullivan_2000, Wilson_2002}. These epochs also see
the bulk of the build-up of the bimodality in galaxy properties of the
local universe, which is apparent in their color distribution
\cite{Baldry_2004}, morphologies \citep{Kauffmann_2004}, spectral
class \citep{Madgwick_2002} and spatial distribution
\citep{Budavari_2003}. Understanding the full picture is complex as
this evolution is the result of the interplay of several physical
processes \citep{Faber_2005} and combine the effects of initial galaxy
formation conditions (``nature'') with galaxy evolution events
(``nurture'') \citep{Kauffmann_2004}. In this context, tracers that
measure over cosmic time galaxy populations selected with homogeneous
physical criteria are of primary interest. They help compare
observations to simulation predictions over a large range of redshifts
with reduced uncertainties, and allow a study of the redshift
evolution of galaxy properties derived from different surveys.

The ultraviolet (UV) range of the spectrum meets these conditions: UV
luminosities provide a good measure of recent star formation within
galaxies \citep{Kennicutt_1998}, modulo attenuation by dust, and has
been widely used at high redshifts to study the properties of the
Lyman Break Galaxies (LBGs) \citep{Giavalisco_2001,
Shapley_2003,Steidel_1995}. As large amounts of data are now becoming
available at lower redshifts as part of the GALEX surveys
\citep{Martin_2005}, the restframe UV spectral domain is presently
well sampled over the full $0<z<6$ redshift range. Furthermore,
comparison of results from high and low $z$ UV-selected samples is
eased by the fact that the UV luminosity density fractions\footnote{We
define the UV luminosity density fraction of a given sample as the
ratio of the UV luminosity density encompassed by the sample over the
total UV luminosity density at the same redshift.} probed at high and
low $z$ are similar \citep[][hereafter Paper II]{Heinis_2007}, due to
the strong luminosity evolution of the UV luminosity function
\citep{Arnouts_2005}. Noticeably, during the epochs probed by GALEX
the properties of active star forming galaxies show a very fast
evolution.

The wealth of UV-selected data now available at low redshifts enables
statistical studies in the context of the downsizing of star
formation, and in particular searches for links between the star
formation properties and galaxy environment in terms of galaxy or dark
matter density. Here we focus on the evolution with redshift of the
link of star formation with Dark Matter and particularly the evolution
of the class of Dark Matter halos hosting actively star forming
galaxies since $z\sim1$. This can be achieved by the study of the
clustering of galaxies: at high redshift, LBGs studies show that UV
selected galaxies inhabit high galaxy density regions
\citep{Steidel_1998}, and are strongly biased with respect to the
underlying Dark Matter, with more actively star forming galaxies being
more biased \citep[][see \citet{Giavalisco_2002} for a review on the
properties of LBGs]{Adelberger_2005, Giavalisco_2001,
Foucaud_2003}. We propose to extend such studies to low redshifts
using similar selection criteria. This paper is the first in a series
and presents the methods and first results of angular clustering
measurements from GALEX data. The following section presents the
datasets and the derivation of the redshift distributions. Section
\ref{sec_acf_methods} presents two methods to derive the angular
correlation function from a set of fields, and a discussion about the
behavior of these methods with respect to photometry inhomogeneity. In
section \ref{sec_acf} we present our results on the angular
correlation functions and correlation lengths. To provide the crucial
link to Dark Matter halos, we use the analytical \citet{Mo_2002}
formalism that we present in sec. \ref{sec_mo_white}. We end with a
short discussion in sec. \ref{sec_discussion}.

Throughout the paper a ${\Lambda}CDM$ cosmology is assumed with matter
density $\Omega_m = 0.3$, vacuum energy density $\Omega_\Lambda=0.7$,
and a Hubble parameter $h=0.7$ where $H_0 = 70 $km s$^{-1}$
Mpc$^{-1}$. All correlation length values taken from the literature
have been converted accordingly using equation (4) in
\citet{Magliocchetti_2000}.

\section{Primary fields selection}\label{sec_selection} 
We use Medium Imaging Survey (MIS) fields from the \galex Release 2
(GR2), which allows us to probe the clustering of faint sources in
\FUV and \NUV at three limiting apparent AB magnitudes 22.0, 21.5 and
21.0. The magnitudes we refer to are corrected for Galactic extinction
using Schlegel maps \citep{Schlegel_1998} and the
\citet{Cardelli_1989} extinction law, unless specified otherwise. The
average color excess in the fields, derived from the
\citeauthor{Schlegel_1998} maps, ranges from $0.08$ to $0.12$. The
extinction coefficients $A_{\FUV}/E(B-V)$ and $A_{\NUV}/E(B-V)$ are
respectively 8.29 and 8.61.

We start with 348 MIS \galex fields overlapping SDSS DR5 of which only
a subset will be kept because of galactic extinction (see
sec. \ref{sec_systematics}). We only include sources within 0.5 $\deg$
radius from the field center, since artifacts concentrate near the
periphery of the field of view \citep[see][]{Morrissey_2005} and
photometric accuracy decreases beyond this limit. We used only objects
within the \galex primary resolution. We used SDSS masks to mask out
holes, bright stars and trails; we used also \galex masks, as well as
additional ones to mask out resolved galaxies or artifacts not
predicted by the \galex pipeline. Only objects with an SDSS match
within $4\arcsec$ are kept, and the closest match is used. Galaxies
are assumed to be SDSS galaxies (\type~= 3 following the morphological
classification of \citet{Lupton_2001} and \citet{Stoughton_2002}). To
check the effects of possible residual QSOs in our sample, we removed
from our sample AGN-dominated objects as objects classified QSOs by a
template fitting procedure\footnote{Le Phare: available and documented
at this URL: {\tt http://www.oamp.fr/arnouts/LE\_PHARE.html}}. Impacts
on the results are found negligible, hence we do not remove QSOs
classified objects from the sample in the following.

\subsection{Redshifts distributions }\label{sec_zdist}
To get the redshift distributions of the samples for the different
magnitude cuts, we use the polynomial fit method described in
\citet{Connolly_1995, Connolly_1997}. This method requires first to be
trained with a spectroscopic sample. We train on 6 bands (\NUV from
\galex, as all objects do not have \FUV photometry, and the 5 SDSS
bands) with 17,843 objects from the sample described in
sec. \ref{sec_selection} having SDSS spectroscopic redshifts. We then
apply the coefficients derived from the training set to the whole
sample.

We performed a simple correction for the broadening due to photometric
redshifts errors by assuming that the photometric redshifts errors
follow a normal distribution independent of the object magnitudes and
redshifts, with the standard deviation $\sigma=0.03$ measured using
all the available spectroscopic redshifts from the SDSS. We check that
the standard deviation does not vary with apparent magnitude using our
photometric redshift estimation on \galex fields with SDSS overlap and
the independent and deeper spectroscopy from \citet{Papovich_2006}.

\begin{figure}
  \plotone{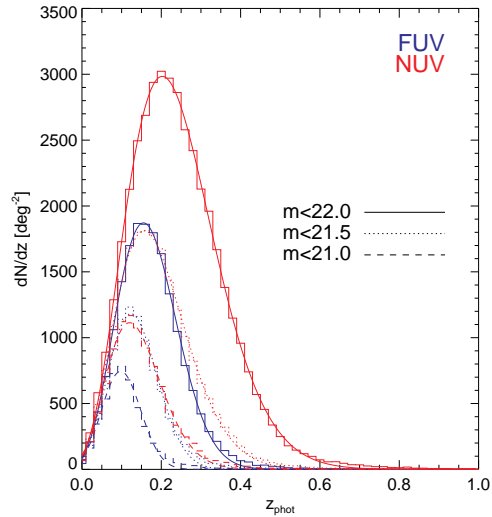}
  \caption{ \small Derived redshift distributions for the two
    \galex bands using three magnitude cuts (histograms). The solid
    curves show the best fitting $N(z)$ convolved by a Gaussian with
    $\sigma=0.03$ (see text).\label{fig_z_dist} }
\end{figure}

Following \citet{Efstathiou_1991}, the parent distribution of the true
redshifts is described by the following parametric shape:
\begin{equation}
  \frac{dN}{dz} = A_z\left(\frac{z}{z_c}\right)^2
  \exp\left[-\left(\frac{z}{z_c}\right)^n\right]
\end{equation}
We fit this shape convolved by a Gaussian with $\sigma=0.03$ to the
observed photometric redshift distribution. Fig.\ref{fig_z_dist} shows
the gaussian-convolved best estimate $N(z)$, along with the measured
distribution for the 174 least extinct fields (see
sec. \ref{sec_systematics}), and table~\ref{tab_Table1} lists the
parameters of the true distributions.

\section{Choice of ACF estimation method}\label{sec_acf_methods}

\subsection{Methods description}

Given that the \galex data are extracted from relatively large numbers
of similar exposures, the question arises as to the most appropriate
method to retrieve the available information. A first, straightforward
approach to measure the ACF from a group of 
non-overlapping
fields is to treat all of them as disjoint subfields of one large,
discontinuous field and to apply the \citet{Landy_1993} (hereafter
LS93) estimator on it:
\begin{equation} 
  w_{CF} = \frac{DD -2DR + RR}{RR}
\end {equation}
where DD (resp. DR, RR) is the number of data-data (resp. data-random,
random-random) pairs from all fields (including cross pairs from
different subfields), normalized by the suitable pair numbers. In the
case of this composite field method (hereafter CF), the number of
random points is fixed for the global field, and not for each
individual \galex field. This is the ideal method which in principle
allows one to extract all the available information. In particular,
this method reduces the integral constraint bias and the noise,
especially at large angular separations.

\begin{figure}
  \plotone{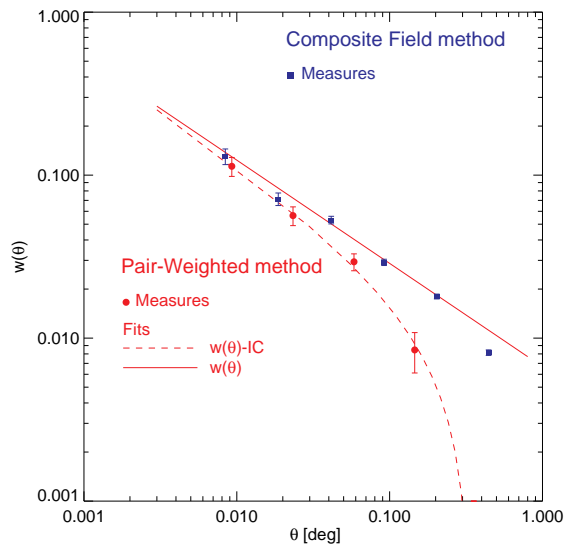}
  \caption{\small Validation of the method used to correct for the
    Integral Constraint bias. The ACF is computed from 50 randomly chosen
    fields in a synthetic catalog with the CF (filled squares) and the
    PW (filled circles) methods. The dashed (resp. solid) line shows
    the best-fit of the PW result uncorrected (resp. corrected) for
    the Integral Constraint bias (see text for details).
\label{fig_IC} }
\end{figure}

Although best in the ideal case, the CF method requires precise
homogeneity of the data and may not be robust in practice.  We
therefore introduce another estimator, which we define as the
following pair-weighted average (hereafter PW) of the ACF measured in
each field individually:

\begin{equation}\label{wcf}
  w_{PW}(\theta) = \frac{\sum_i \widetilde{RR_i}(\theta) w_i(\theta)}
  {\sum_i \widetilde{RR_i}(\theta)}
\end{equation} 
where $w_i$ is the ACF estimated from field $i$ alone computed with
the LS93 estimator, and $\widetilde{RR_i}$ the number of random-random
pairs in the random catalogue constructed for this field (see Appendix
\ref{app_pw} for a derivation of this formula (eq .\ref{wcf}) from
$w_{CF}$). The $\widetilde{RR_i}$ term involves pair numbers and field
geometry information. The PW method is by construction insensitive to
field-to-field fluctuations -- and thus best suited for the peculiar
MIS geometry. A drawback of the PW method is the increase of the
integral constraint (IC) bias because of the smaller angular extent of
the field\footnote{In the case of independent fields, the IC in the PW
method is typically higher than that of the CF method by a factor of
the number of fields.} as well as an increase in the noise. The
integral constraint can be relatively well corrected for using its
estimate given by LS93. To compute it, we assume that the real
correlation function is a power-law $A_w\theta^{-\delta}$ and we fit
$A_w\theta^{-\delta} - I(A_w,\delta)$ to the data, where
$I(A_w,\delta) = 1/\Omega^2 \int_{\Omega} A_w \theta^{-\delta}
d\Omega_1 d\Omega_2$, integrated over a \galex field. This method is
similar to that used by \citet{Roche_1999}, except that $\delta$ is
left as a free parameter. In the following, ``PW method'' will refer
to the IC-corrected technique. We have checked the accuracy of the
above correction of the IC bias using a $100 deg^2$ synthetic catalog
derived from \verb|GalICS| \citep{Hatton_2003, Blaizot_2005}. The ACF
has been computed with the CF and the PW methods from 50 randomly
positioned fields of radius $0.5^{\circ}$. The \verb|GalICS|-magnitude
cut was chosen to obtain approximately the same mean number of
galaxies as found using the \NUV$<22$ cut. The results of the CF and
PW methods have been found undistinguishable for the model catalogs
(fig \ref{fig_IC}).

\subsection{Systematic effects}\label{sec_systematics}

\begin{figure*}
  \includegraphics[width=\hsize]{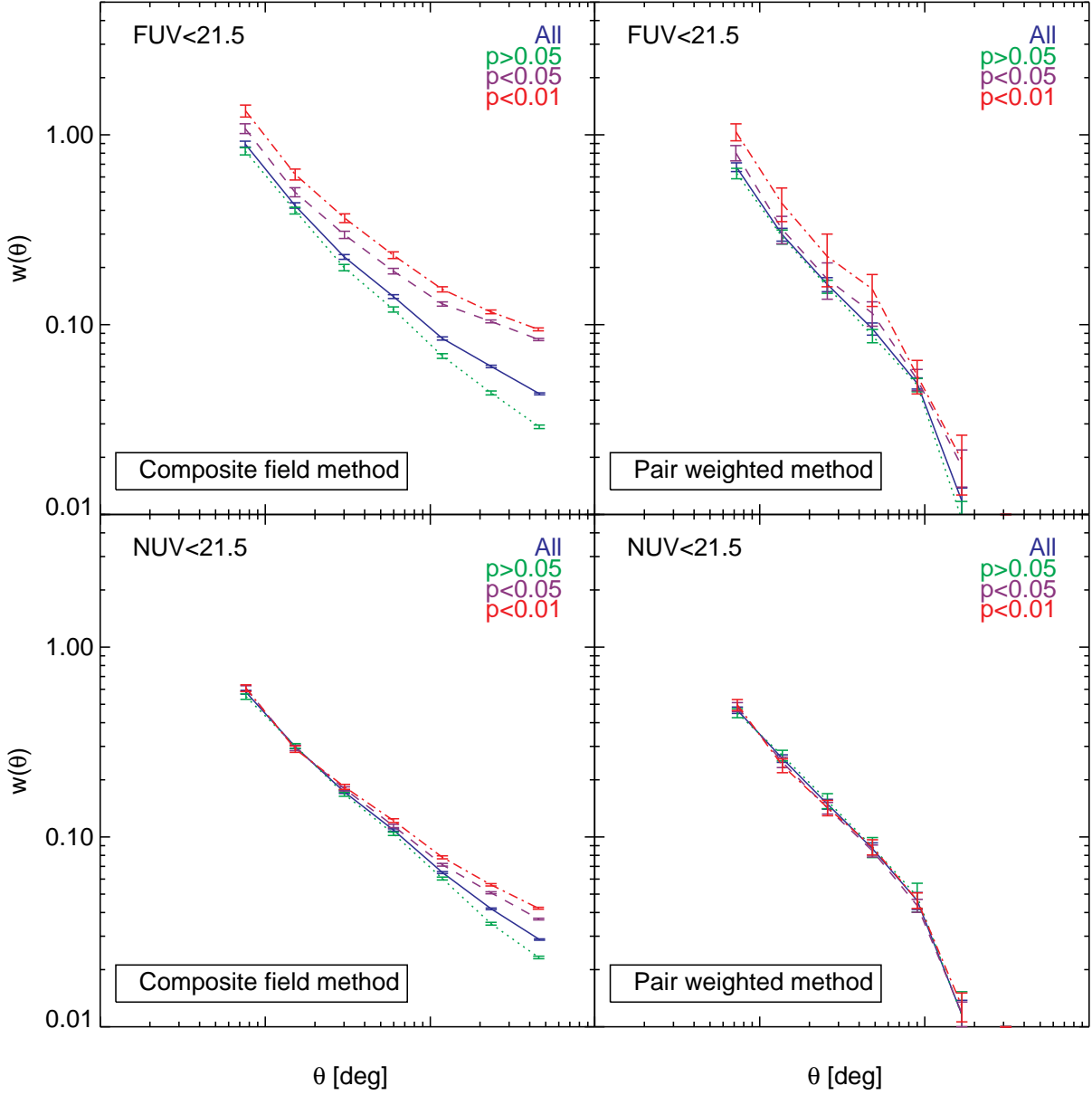}
  \caption{ \small Angular correlation function for the CF (left) and
    PW (right) methods (see text). Upper panels show $\FUV <21.5$
    selection and lower panels $\NUV <21.5$. The solid line shows the
    ACF for the whole sample; the other curves represent the ACF of
    the fields according to the probability that their magnitude
    distribution is drawn from the sample distribution than the whole
    sample ($p<0.01$, dot-dashed; $p<0.05$, dashed; $p>0.05$, dotted.)
    \label{fig_dust_extinction} }
\end{figure*}

To test the sensitivity of the CF and PW methods to systematics, we
used a statistical approach to decide whether the photometry of a
given field is drawn from the same distribution than the photometry of
the whole sample. To this aim, we used the Mann-Whitney test, which is
independent of the size of the input samples; we use only objects with
\type = 3 during this process to avoid strong star contamination. For
each field, we create a test sample built from the magnitudes of the
objects in this field, and a control sample from the magnitudes of the
objects that belong to all other fields. The Mann-Whitney test
provides as output the probability that these two distributions are
the same. 
We show figure \ref{fig_dust_extinction} the ACFs of fields
grouped according to their value of this probability, using the CF
method (left), or the PW method (right), for $\FUV <21.5$ (top) or
$\NUV <21.5$ (bottom). The results obtained from the CF and PW methods
show significant differences. The CF method results show an excess of
power especially at large scales; the amplitude of this excess
increases as the probability that the field photometry is the same
than the overall sample photometry decreases. Conversely the results
of the PW method are fairly insensitive to photometry inhomogeneities;
there is an overall power excess at $\FUV <21.5$ for the fields with a
probability lower than 0.01, but the ACF of the whole sample is very
similar to the ACF of the best fields ($p>0.05$).

There are several sources of systematic errors, which, although their
individual effects are weak, may, combined with each other, yield the
trends observed. Similar trends are observed when the fields are
binned according to the mean Galactic extinction. However, the
cross-correlation between galaxies and dust maps using both CF and PW
methods is found at least 5 times lower than the autocorrelation at
scales where the latter is positive ($\theta\lesssim 0.2\deg$), and no
obvious trend was found between the amplitudes of this
cross-correlation and the Galactic extinction. On the other hand, the
amplitude of the cross-correlation function between galaxies and
background maps is higher in fields with higher mean Galactic
extinction. Inhomogeneities may also arise from photometry drift with
time, but sources drifted less than 0.1 magnitude
\citep{Morrissey_2007} over the whole GALEX mission; a drift of this
amplitude has a small effect on the CF method, as expected from tests
on mock catalogs. Studying sources observed several times in
overlapping regions shows that field-to-field fluctuations are less
than 10 \% beyond what is expected from Poisson statistics. Note
however that this result is based on a few sources per field located
at the edges of the field, where photometry accuracy decreases. Star
contamination can lower the amplitude of the PW method, as an addition
of an uncorrelated population, while it may contribute to the effects
observed with the CF method, given the variations of star counts with
Galactic latitude. According to template-fitting based classification,
the fraction of stars in SDSS objects with \type~= 3 is 2 \% in NUV
and 8 \% in FUV; we checked that this has small effect on the PW
method.

All these tests suggest that there is some source of field-to-field
variations in our data, likely due to a combination of zero-point
calibrations, background fluctuations (correlated with Galactic
extinction), etc ... The PW method is fairly insensitive to any
systematics, as expected, and we are thus confident that it is a
robust estimator. It is this method we chose to use in the rest of the
paper.

Conservatively, for the remainder of the paper, we restrict the
analysis to the 174 fields with the lowest Galactic extinction
($\langle E(B-V) \le 0.04 \rangle$).  The number of galaxies at the
different limiting magnitude cuts are given in table~\ref{tab_Table1}.

The characteristics of the UV dust attenuation in galaxies are not
known to be correlated with the large scale structure or the galaxy
density \citep[even in the extreme cases of clusters, see
e.g.][]{Boselli_2006}; the effect of internal dust has thus been taken
as an uncorrelated noise source on the UV fluxes and its effect on the
ACF neglected. This allows direct comparison with clustering studies
of high redshift restframe UV-selected galaxies.

\begin{deluxetable}{cccccccc}
\tablecolumns{6} \tabletypesize{\footnotesize} \tablewidth{0pt}
\tablecaption{Sample description, power-law best fits parameters and
  comoving correlation lengths \label{tab_Table1}} 
\tablehead{
\colhead{}    &  \multicolumn{3}{c}{FUV} &   \colhead{}   &
\multicolumn{3}{c}{NUV} \\
\cline{2-4} \cline{6-8} \\ 
\colhead{Limiting mag} & \colhead{22.}   & \colhead{21.5}    & \colhead{21.}  &
\colhead{}    & \colhead{22.}   & \colhead{21.5}    & \colhead{21.}  }
\startdata
$N_{gal}$\tablenotemark{*}            & 44651   & 22655   & 11418   & & 99368  & 48274  & 22948   \\[0.1cm]
$\overline{z}$\tablenotemark{\dagger} & 0.18    & 0.15    & 0.12    & & 0.25   & 0.21   & 0.16    \\[0.1cm]
$A_z$\tablenotemark{\ddagger}         & 5485.96 & 3616.28 & 2417.72 & & 8255.5 & 5084.1 & 3291.14 \\[0.1cm]
$z_c$\tablenotemark{\ddagger}         & 0.16    & 0.13    & 0.11    & & 0.18   & 0.13   & 0.1     \\[0.1cm]
$n$\tablenotemark{\ddagger}           & 2.18    & 2.29    & 2.76    & & 1.67   & 1.61   & 1.66    \\ [0.1cm]
$A_{w}\times10^3$ $\left[\deg^{\delta}\right]$ & 8.0$^{+1.7}_{-1.5}$ & 9.7$^{+3.7}_{2.7}$ & 10.4$^{+7.2}_{-4.4}$  & & 3.2$^{+0.5}_{-0.5}$ &4.6$^{+1.1}_{-0.8}$ & 6.3$^{+0.9}_{-0.9}$  \\[0.1cm]
$\delta$      & 0.80$\pm$0.05 & 0.80$\pm0.08$ & 0.75$\pm0.12$        & & 0.89$\pm0.04$ & 0.88$\pm0.05$ & 0.84$\pm0.09$   \\[0.1cm]
$r_0$ [Mpc]   & 4.2$^{+0.5}_{-0.4}$ & 3.7$^{+0.7}_{-0.6}$ & 2.8$^{+0.9}_{-0.7}$  & & 4.0$^{+0.3}_{-0.3}$ & 3.7$^{+0.4}_{-0.3}$& 3.3$^{+0.7}_{-0.5}$ \\[0.1cm]
\enddata

\tablenotetext{*}{Number of galaxies in the samples}
\tablenotetext{\dagger}{Mean photometric redshift}
\tablenotetext{\ddagger}{Parameters of the true best fit redshift
distribution (see text)} \tablecomments{The amplitude and slope of
best fit power laws to the angular correlation function, and hence the
comoving correlation length account for the Integral Constraint
correction (see text). No attempt to remove residual QSOs from
photometric redshifts is performed here.}
\end{deluxetable}

\section{Angular correlation function and correlation length}\label{sec_acf}

\subsection{Measurements}

\begin{figure*}[t]
  \includegraphics[width=\hsize]{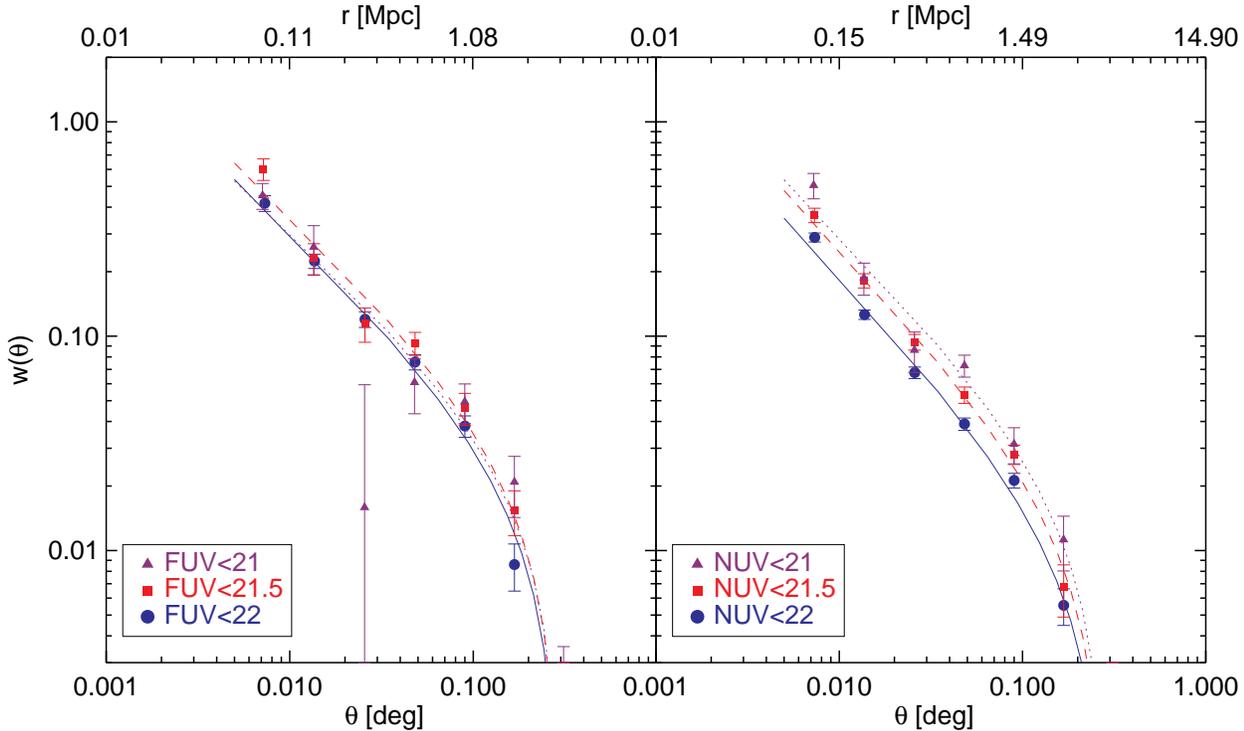}
  \caption{ \small Angular correlation function measured in the 174
    \galex fields with the lowest Galactic extinction, for the 
    \FUV (left) and \NUV (right) \galex bands at three magnitude cuts:
    circles, $m_{UV}<22$; squares, $m_{UV}<21.5$ and triangles,
    $m_{UV}<21$. Dashed lines show the power law best-fit uncorrected
    for Integral Constraint. The upper axis shows the comoving
    distances corresponding to the angular scales at $z=0.15$. No
    attempt to remove residual active nuclei by photometric redshift
    template fitting has been made beyond the SDSS
    classification. \label{fig_ACF} }
\end{figure*}

We use the PW method described in sec. \ref{sec_acf_methods} to
measure $\omega(\theta)$ from the 174 fields with the lowest Galactic
extinction, using logarithmic-width bins of $\Delta\log\theta = 0.27$,
and $\theta_{min}=0.005\degr$ and $\theta_{max}=0.4\degr$, which
probes scales in the range 0.05 to 4 comoving Mpc at the median
redshift ($z=0.15$) of the samples considered here. The results are
plotted in fig.~\ref{fig_ACF}; the $1\sigma$ errorbars represent
internal scatter derived from jackknife resampling of the 174 \galex
fields used for the ACF. In order to check for any instrumental
contribution to the ACF such as residual non uniformities of the
sensitivity across the field of view, the PW method has been applied
to stars, selected as objects whose SDSS counterparts with \type~=
6. For stars we find no significant deviation from a null correlation
function.

We fit the results using the method described
in sec. \ref{sec_acf_methods}. Our best fits for the different samples
are given in table \ref{tab_Table1}, where $\theta$ is expressed in
degrees. The error bars on $A_w$ and $\delta$ are the projected
$(\chi^2_{min}+1)$ contour.

\begin{figure}[t]
\includegraphics[width=\hsize]{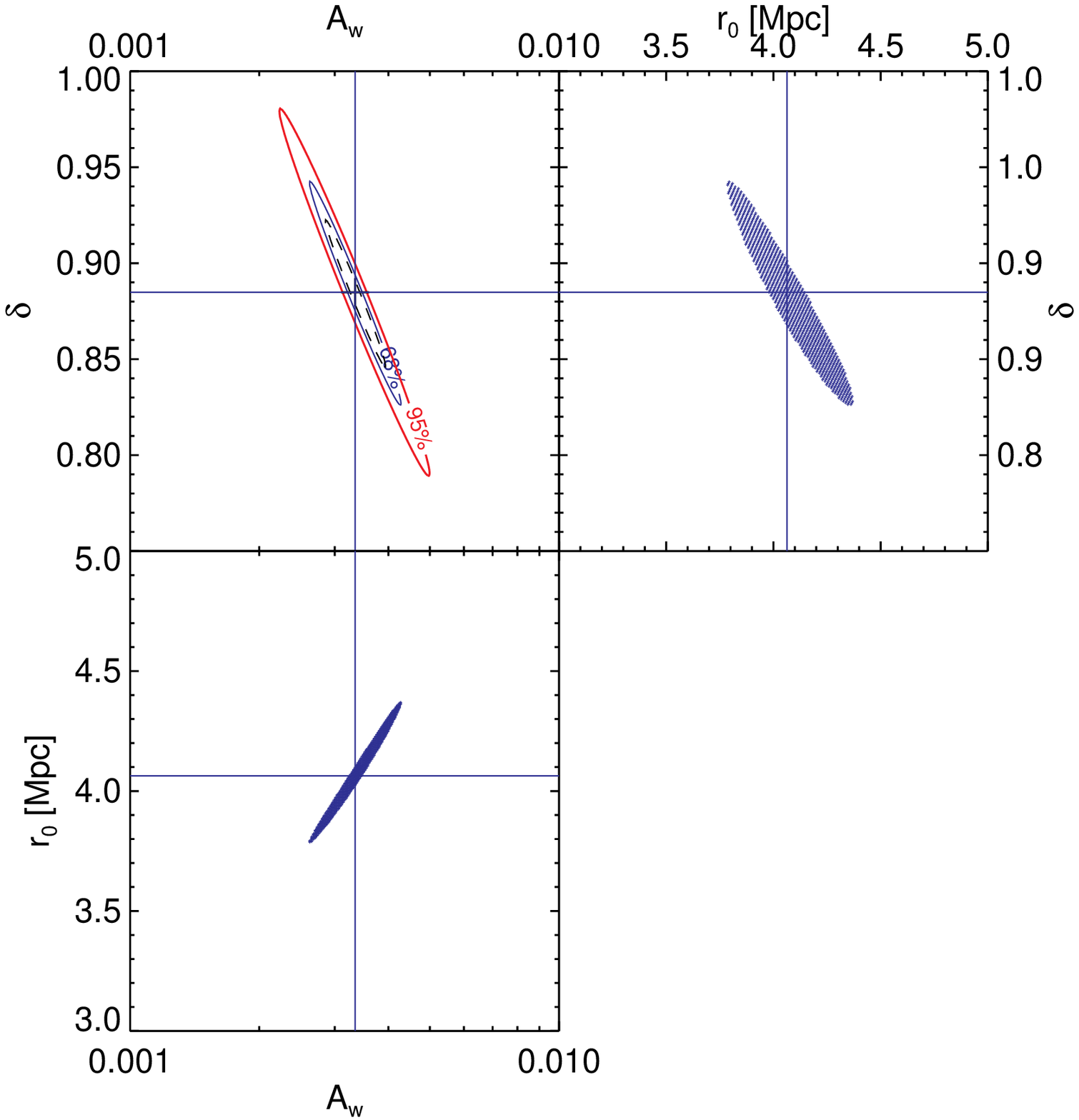}
  \caption{ \small $\chi^2$ contours and derivations of errors bars on
  $A_w$, $\delta$ and $r_0$ for the \NUV$<22$ result. \textit{Top
  left}: contours of constant $\chi^2$ in the $(A_w,\delta)$
  plane. The inner (resp. outer) solid line corresponds to the 68.3\%
  (resp. 95.4\%) confidence level. The dashed line shows the
  $\chi_{min}^2+1$ contour; its projections on the axes give the error
  bars on $A_w$ and $\delta$. \textit{Top right}: comoving $r_0$ as a function
  of $\delta$ given $A_w$ using the values of $A_w$ and $\delta$
  included in the 68.3\% confidence level. \textit{Bottom left}: comoving $r_0$
  as a function of $A_w$ given $\delta$ using the same values. The
  errors bars on $r_0$ are the extrema of this distribution. The solid
  line intersections show the location of the best fit in each panel.
  \label{fig_chi2}}
\end{figure}

To derive the comoving correlation length, $r_0$, we used the Limber
equation \citep{Peebles_1980} with the true deconvolved redshift distributions
(see in sec. \ref{sec_zdist}). The results are given in table
\ref{tab_Table1}. The uncertainties on $r_0$ have been assumed to be
the extreme excursions of $r_0$ in the projection in the
($r_0,\delta$) and ($r_0,A_w$) planes of the $\chi^2$ contour at the
$68\%$ probability in the ($A_w,\delta$) plane (see fig
\ref{fig_chi2}).

\subsection{Comparison with previous studies}
Given the error bars, the slopes $\delta$ found for the different
magnitude cuts in the two bands are compatible with a constant value
$\delta\simeq0.81\pm0.07$. This is steeper than reported in several
studies based on blue galaxies at low $z$ \citep{Budavari_2003,
Zehavi_2002, Madgwick_2003}, and restframe UV-selected galaxies at
higher redshifts \citep{Adelberger_2005, Porciani_2002}, all of them
consistent with a value of $\delta\simeq0.6$. However our measurement
is in agreement with \citet{Giavalisco_1998, Giavalisco_2001,
Foucaud_2003}. Moreover, \citet{Coil_2004} noticed a steepening of the
slope not only for the reddest but also for the bluest galaxies of
their samples. We discuss in Paper II the dependence of $\delta$ on UV
luminosity.

With an average comoving correlation length $3.7\pm0.6$ Mpc at $z \sim
0.2$, the present \galex data sets confirm the low clustering of the
rest-UV selected galaxies in the local universe observed by
\citet{Heinis_2004}. The new mean value is 25\% lower, though
both measurements agree within error bars. Assuming the average values
of $r_0$ and $\delta$ quoted above, the corresponding bias defined at
8 Mpc by ${b_8}=\sigma_{8,g}/\sigma_{8,m}$
\citep[e.g.][]{Magliocchetti_2000} is $0.61 \pm 0.09$ at $z=0.2$, a
significant anti-bias.

It is well-known that blue galaxies are less strongly correlated than
red ones, and not surprisingly the small correlation length found in
this study is comparable to that measured for blue galaxies in the
local universe: \citet{Coil_2004} report a comoving $r_0$ of
$2.54\pm0.37$~Mpc for the class of blue galaxies defined by ($0.2 <
R-I< 0.4$) in their visible-selected sample, which spans the redshift
range $0.3-0.6$, with the lowest correlation length among all of their
galaxy subsamples. The \galex restframe UV selected galaxies are
nevertheless even less correlated than the galaxy class T4 (bluest of
4 classes) from \citet{Budavari_2003} for which they derive a $r_0$ of
$6.44\pm0.27$~Mpc. \citet{Hawkins_2001} computed the redshift-space
correlation function from far-infrared selected galaxies in the local
Universe ($z \sim 0.03$). Converted to real-space, their estimate of
the correlation length ($r_0 = 5 \pm 0.33$ Mpc) of the hotter
galaxies, i.e. the most star-forming, is higher than ours from
UV-selected galaxies.

Low-redshift restframe UV-selected galaxies possess correlation
lengths slightly lower than those derived from high-$z$ restframe
UV-selected samples (see fig. \ref{fig_ro_halos_z}). Note that the
comparison with results from higher $z$ samples is not straightforward
because of the UV luminosity segregation \citep{Giavalisco_2001,
Adelberger_2005, Zehavi_2005, Norberg_2002}: brighter objects are more
clustered than fainter ones. The \galex samples are the faintest of
the restframe UV-selected samples considered here: the mean absolute
magnitudes of the \FUV~and \NUV~samples are $M_{\FUV}=-18.3$ and
$M_{\NUV}=-18.8$, while the Lyman break galaxies samples of
\citet{Adelberger_2005, Arnouts_2002, Foucaud_2003, Giavalisco_2001}
are all brighter than $M_{UV}=-20$. We study in details the luminosity
dependence of clustering within the \galex samples in Paper II.

\subsection{Comparison with Dark Matter Halo clustering predictions}\label{sec_mo_white}

\begin{figure}[t]
  \includegraphics[width=\hsize]{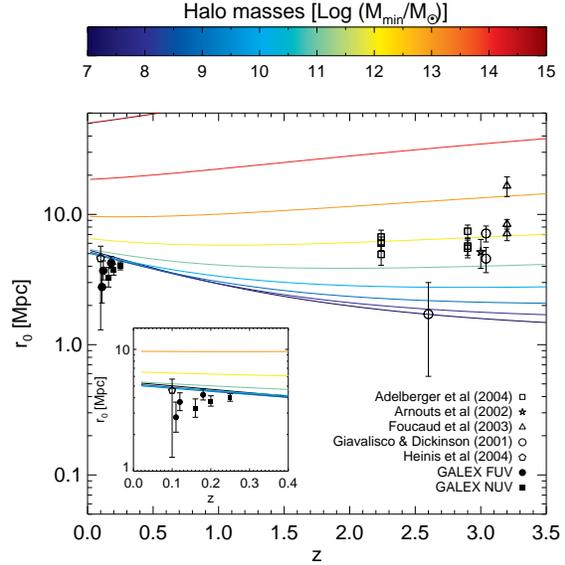}
  \caption{ \small Comparison of the evolution with redshift of the
  correlation lengths obtained from restframe UV selected samples with
  the correlation lengths of Dark Matter Halos more massive than
  $M_{min}$ (color-coded). The inset shows the low $z$ points.
  \label{fig_ro_halos_z}}
\end{figure}

In this section we use the formalism described by \citet{Mo_2002} to
compute the correlation length of Dark Matter Halos (DMHs) above a
given mass as a function of redshift. We assume that the spatial
correlation function of DMHs with masses greater than $M_{min}$ at a
redshift $z$ is well approximated by a power-law:
\begin{equation}
  \xi(r,M_{min},z) = \left(\frac{r}{r_{0_h}(M_{min},z)}\right)^{-\gamma_h}
\end{equation}
where $r_{0_h}(M_{min},z)$ is the correlation length of such
halos. The \citet{Mo_2002} formalism provides analytical equations for the
abundance and the bias factor of the halos, $n(M,z)$ and $b(M,z)$,
given their mass and redshift. The effective bias of the halos more massive
than a minimum mass $M_{min}$ at a redshift $z$ is then given by:
\begin{equation}
  b_{eff}(M_{min},z) = \frac{\int_{M_{min}}^{\infty}
  b(M,z)n(M,z)dM}{\int_{M_{min}}^{\infty}n(M,z)dM}
\end{equation}
The rms density fluctuations of the halos is linked to the rms density
fluctuations of the underlying mass at 8 $h^{-1}$Mpc by:

$\sigma_{8,h}(M_{min},z) = b_{eff}(M_{min},z) \sigma(z)_{8,m}$

where the subscripts $h$ and $m$ denote respectively halos and underlying mass,
and $\sigma(z)_{8,m} = \sigma(0)_{8,m}D(z)$
\citep[see][]{Mo_2002} with $\sigma(0)_{8,m} = 0.9$. The correlation
length of the DMHs with masses $M>M_{Min}$ at $z$ is then obtained
using \citep[e.g.][]{Magliocchetti_2000}:
\begin{equation}
  \sigma_{8,h}(M_{min},z) = \sqrt{C_{\gamma_h}\left(
  \frac{r_{0_h}(M_{min},z)}{8}\right)^{\gamma_h}}
\end{equation}
where $C_{\gamma} = 72/[(3-\gamma)(4-\gamma)(6-\gamma)2^{\gamma}]$. We
assumed that the slope of the spatial correlation function of the
halos is $\gamma_h = 1.8$, after having checked that the results are
rather insensitive to the adopted value if $1.5<\gamma_h<2.5$. Fig.~
\ref{fig_ro_halos_z} shows the redshift evolution of the correlation
length of DMH with masses $10^7<M_{min}<10^{15} M_{\odot}$.

In the framework of Halo Occupation Distribution (HOD) models
\citep[e.g.][]{Berlind_2002, Cooray_2002}, recent studies have pointed
out that the galaxy correlation function is likely to be the sum of
two components.  The first component dominates at small scales,
describing the correlation of galaxies that are in the same halo, and
the second component accounts for galaxies in different halos,
dominating at large scales. The present sample however does not
provide sufficient constraint to fit HOD models to our results---we
plan to perform this in future papers. Assuming that the correlation
function of DMHs is a power-law at all scales and that each halo hosts
at most one galaxy, a direct comparison of the \galex results with the
correlation lengths of dark matter halos (fig~\ref{fig_ro_halos_z})
shows that the UV selected galaxies in our samples have the same
correlation lengths as halos with masses lower than $M_{min} =
10^{11}M_{\odot}$. At $z>2$, clustering measurements from LBGs samples
show that halos with comparable clustering strengths have $M_{min}
\gtrsim 10^{12}M_{\odot}$, as already mentioned by
\citet{Adelberger_2005} and \citet{Giavalisco_2001}. These results
suggest that the characteristic mass of halos hosting active star
formation has decreased from $z = 3$. Note that taking luminosity
evolution into account does not weaken this result, since at low
redshifts UV-selected samples actually probe the same UV LD fraction
as their high redshift counterparts (see Paper II). This mass
evolution, as well as the bias evolution can be interpreted as
additional evidence for the ``downsizing'' scenario of star formation,
though it applies here to the mass of the underlying halo rather than
to the baryonic mass. Note that these is a correlation between halo
and galaxy mass \citep{Shankar_2006}, though its scatter is expected
to be stronger for star forming galaxies \citep[see
e.g.][]{Yoshikawa_2001}. Theoretical studies also predict that the SFR
increases with halo mass \citep[e.g. at $z = 3$,][]{Bouche_2005};
however, in presence of AGN feedback\citep[e.g.][]{DiMatteo_2005,
Croton_2006}, or when taking into account gravitational heating
\citep{Khochfar_2007}, this trend reverses at lower redshift.

The conclusions of this work are developed in Paper II.

\section{Conclusions}\label{sec_discussion} 

We presented here the first clustering measurements from the \galex
data. These data provide a unique basis to statistical studies of star
formation in galaxies at low redshift from their UV continuum. The
same tracer can now be used in an homogeneous way over a large
redshift range ($0<z<4$) to investigate the processes driving star
formation evolution. We discussed the impact of the Galactic
extinction on the clustering measurements and used a method
insensitive to the color excess. We measured the clustering by the
angular correlation function, and fitted our results with a power law
parametrisation: $w(\theta) = A_w \theta^{-\delta}$. We derive steep
slopes, $\delta \simeq 0.81 \pm 0.07$. Assuming photometric redshift
estimation, we compute the correlation length, $r_0$. The results
confirm the low clustering of UV-selected galaxies at low redshift
($r_0 = 3.7 \pm 0.6$ Mpc). Comparison with analytical modeling shows that
active star forming at $z<0.4$ present the same correlation lengths
than DMHs with $M_{min} < 10^{11} M_{\odot}$. This result is in
agreement with the ``downsizing'' scenario.

\acknowledgments It is with great pleasure that we thank Jean-Michel
Deharveng for support and discussions. \galex (Galaxy Evolution
Explorer) is a NASA Small Explorer, launched in April 2003. We
gratefully acknowledge NASA's support for construction, operation, and
science analysis for the \galex mission, developed in cooperation with
the Centre National d'Etudes Spatiales of France and the Korean
Ministry of Science and Technology.

\appendix
\section*{Pair-Weighted Average estimator for the Angular Correlation Function  }\label{app_pw}
In this section we discuss the weighted estimator presented in sec.
\ref{sec_acf_methods}. The expression can be derived directly from the
definition of the LS93 estimator with some assumptions. The LS93 estimator
is:
\begin{equation}
  w_{LS}(\theta) = \frac{DD(\theta) -2DR(\theta) + RR(\theta)}{RR(\theta)}
\end{equation}

where $DD$, $DR$ and $RR$ are normalized by the suitable pairs number :
\begin{eqnarray}
  DD & = & \frac{2\widetilde{DD}}{n_g(n_g-1)}\\
  DR & = & \frac{\widetilde{DR}}{n_g n_r}\\
  RR & = & \frac{2\widetilde{RR}}{n_r(n_r-1)}
\end{eqnarray}

where $n_g$ is the number of galaxies in the sample and $n_r$ is the
number of random objects in the random sample. In the following we do
not recall the $\theta$ dependence of the different quantities.

Let us consider the case of the CF method (see
sec. \ref{sec_acf_methods}) applied on $N$ fields positioned on the sky
in such a way that no cross pair between objects from different fields
has to be accounted for in the computation of $w(\theta)$. The total
number of pairs over all the fields in each angular bin can then be
expressed using the number of pairs in each field:

\begin{equation}
 \widetilde{DD} = \sum_{i=1}^{N} \widetilde{DD_i} = \sum_{i=1}^{N}
 \frac{n_{g_i}(n_{g_i}-1)}{2} DD_i
\end{equation}
where $\widetilde{DD_i}$ is the number of data-data pairs and
$n_{g_i}$ the number of galaxies in the $i$th field. The same equations
hold for $\widetilde{DR}$ and $\widetilde{RR}$.

When computing the ACF of one field individually, we consider 100
random samples with the same number of random points that galaxies in
this field\footnote{In the case of the CF method, the total number of
random points would also be fixed to $n_g$, but the number of random
points in each field is allowed to be different of $n_{g_i}$.}. $RR_i$
is then the average of the 100 computations. So $n_{g_i} = n_{r_i} =
n_i$ and $n_g = n_r = n$. Then the LS93 estimator can be written:

\begin{equation}\label{eq_start}
  w = \frac{1}{\frac{2\widetilde{RR}} {n(n-1)}} \left[ \frac{2}
  {n(n-1)} (\widetilde{DD} + \widetilde{RR}) -\frac{2}{n^2}
  \widetilde{DR} \right]
\end{equation}

Let us consider the term in brackets; with our assumptions it yields:
\begin{equation}
  \frac{2} {n(n-1)}\sum_i \frac{n_i(n_i-1)}{2} \left(DD_i + RR_i\right) -
  \frac{2}{n^2} \sum_i n_{i}^{2} DR_i
\end{equation}

then introduce the term $RR_i/RR_i$ in both sums:
\begin{equation}\label{eq_interm1}
  \frac{2} {n(n-1)}\sum_i \frac{n_i(n_i-1)}{2} \left(DD_i +
  RR_i\right) \frac{RR_i}{RR_i} - \frac{2}{n^2} \sum_i n_{i}^{2} DR_i
  \frac{RR_i}{RR_i}
\end{equation}

The ACF of the ith field can be written as:
\begin{eqnarray*}
  w_{i}   & = & w_{1_i} + w_{2_i} + w_{3_i} \textrm{ with} \\
  w_{1_i} & = & \frac{DD_i}{RR_i}\\
  w_{2_i} & = & -2\frac{DR_i}{RR_i}\\
  w_{3_i} & = & \frac{RR_i}{RR_i}
\end{eqnarray*}

Hence \ref{eq_interm1} becomes
\begin{equation}\label{eq_interm2}  \frac{2} {n(n-1)}\sum_i \frac{n_i(n_i-1)}{2} RR_i\left( w_{1_i} +
  w_{3_i} \right) + \frac{1}{n^2} \sum_i n_{i}^{2} RR_i w_{2_i}
\end{equation}

At this stage we also assume that $n_i \gg 1$ so that $n_i(n_i-1)
\simeq n_{i}^{2}$, and hence $n(n-1) \simeq n^{2}$; \ref{eq_interm2}
yields

\begin{equation}\label{eq_interm3}
  \frac{2} {n^2}\sum_i \widetilde{RR_i}w_i
\end{equation}

Coming back to eq. \ref{eq_start} we finally get

\begin{equation}\label{eq_end}
  w_{PW}(\theta) = \frac{\sum_i \widetilde{RR_i}(\theta)w_i} {\sum_i
  \widetilde{RR_i}(\theta)}
\end{equation}

\end{document}